\documentclass[aps,pre,twocolumn,superscriptaddress,showpacs]{revtex4-1}
\usepackage[pdftex]{graphicx}
\usepackage{graphicx}
\usepackage{dcolumn}
\usepackage{bm}
\usepackage{amsmath,amssymb}
\usepackage[
  colorlinks=true,
  allcolors=blue,
  breaklinks=true
]{hyperref}

\usepackage[english]{babel}
\usepackage{times}
\usepackage{amsfonts}
\usepackage{psfrag}
\usepackage{verbatim}
\usepackage{color}

\newcommand{\beq}{\begin{equation}}
\newcommand{\eeq}{\end{equation}}
\newcommand{\barr}{\begin{eqnarray}}
\newcommand{\earr}{\end{eqnarray}}
\newcommand{\bei}{\begin{itemize}}
\newcommand{\eei}{\end{itemize}}

\newcommand{\G}{\Gamma}

\begin{document}


\title{Measuring effective temperatures in a generalized Gibbs ensemble}

\author{Laura Foini} 
\affiliation{Department of Quantum Matter Physics, University of Geneva, 24 Quai Ernest-Ansermet, CH-1211 Geneva, Switzerland}
\affiliation{Laboratoire de Physique Statistique (UMR 8550 CNRS), \'Ecole Normale Sup\'erieure ´
Rue Lhomond, 75005 Paris, France}

\author{Andrea Gambassi}
\affiliation{SISSA -- International School for Advanced Studies and INFN, via Bonomea 265, 34136 Trieste, Italy}

\author{Robert Konik}
\affiliation{CMPMS Division, Brookhaven National Laboratory, Building 734, Upton, New York 11973, USA}

\author{Leticia F. Cugliandolo} 
\affiliation{Sorbonne Universit\'es, Universit\'e Pierre et Marie Curie
  -- Paris 6, Laboratoire de Physique Th\'eorique et Hautes Energies,\\
  4, Place Jussieu, Tour 13, 5\`eme \'etage, 75252 Paris Cedex 05,  France}

\received{\today}


\begin{abstract}
The local physical properties of an isolated quantum statistical
system in the stationary state reached long after a quench are
generically described by the Gibbs ensemble, which involves only its
Hamiltonian and the temperature as a parameter. 
If the system is instead integrable, 
additional quantities conserved by the dynamics intervene in the
description of the 
stationary state. 
The resulting generalized Gibbs ensemble involves a number of
temperature-like parameters, the determination of which is
practically difficult. 
Here we argue that in a number of simple models these parameters can
be effectively determined by using fluctuation-dissipation relationships
between response and correlation functions of natural observables,
quantities which are accessible in experiments. 
\end{abstract}

\pacs{PACS}

\maketitle

\setlength{\textfloatsep}{10pt} 
\setlength{\intextsep}{10pt}


\section{Introduction}

A question of fundamental relevance in the physics of many-body quantum systems is under which conditions and in which sense they can reach an eventual equilibrium state after evolving in isolation from the surrounding environment.   
With this issue is mind, a considerable theoretical and experimental effort is currently devoted  to studying 
the dynamics of these systems 
from various 
perspectives~\cite{Polkovnikov10,Gogolin,Pasquale-ed}. 

A statistical system reaches a steady state if the 
long-time limit of the reduced density matrix of any of its subsystems 
exists. This means that, given a generic local observable within the latter,
its expectation value  on the (unitary) evolution of the initial state $|\psi_0\rangle$ can be alternatively determined as a statistical average over an ensemble with density matrix $\hat \rho$. Accordingly, local multi-time correlation functions within a subsystem can be determined 
either from the dynamics or, equivalently, as averages over $\hat \rho$.

In an isolated  non-integrable system with Hamiltonian $\hat H$, 
the statistical density matrix $\hat \rho$ is the familiar 
Gibbs-Boltzmann one, i.e., 
$\hat \rho = Z^{-1} e^{-\beta \hat H}$
in which the inverse temperature $\beta$ is fixed by the 
conservation of energy, $\langle \psi_0 | \hat H | \psi_0\rangle = {\rm Tr} ( \hat \rho \hat H)$ and $Z$ is the partition function. 
In an integrable system, instead, the needed statistical ensemble is
expected to 
be the so-called generalized Gibbs ensemble (GGE)~\cite{Rigol08} $\hat \rho_{\rm GGE}$, 
which, as detailed further below in Sec.~\ref{sec:GGE-FDR},
involves the set of mutually commuting, (quasi-)local \cite{Prosen} charges $\hat Q_k$ which are conserved by the dynamics  and the associated parameters 
$\lambda_k$ --- a sort of 
inverse temperatures --- with as many elements as necessary. 
Although in principle feasible, constructing
$\hat \rho_{\rm GGE}$ can be notably difficult, as identifying all the $\hat Q_k$'s~\cite{Illievski15} is not always simple. 
Once this is done, the values of the parameters $\lambda_k$ can be 
determined as one does for the temperature in the canonical ensemble, i.e., by requiring that the expectation value of each $\hat Q_k$ calculated on $\hat \rho_{\rm GGE}$ matches the (conserved) initial value $\langle \psi_0| \hat Q_k |\psi_0\rangle$. In this respect, the $\lambda_k$'s are nothing but 
the Lagrange multipliers used to enforce a set of constraints.  
It was stressed in Ref.~\cite{CauxKonik} that these 
multipliers appear also in the generalized
thermodynamic Bethe ansatz analysis of a class of models, not individually, but in certain combinations.  
However, neither viewpoint on the $\lambda_k$'s
provides a straightforward means to characterize them in experiment.

In practice, the values of the $\lambda_k$'s can be inferred by fitting 
various other quantities with the theoretical predictions stemming from 
the corresponding theoretical model in its asymptotic state described by $\hat \rho_{\rm GGE}$.
This route was followed in 
Ref.~\cite{GGE-exp}, in which the equal-time spatial correlation functions of the relative phase between two halves of a one-dimensional 
Bose gas after splitting are compared, during relaxation, with the predictions of the low-energy Luttinger liquid description of the corresponding 
Lieb-Liniger model. The ten most relevant $\lambda_k$'s were thus determined as fitting parameters.
However, due to its indirect nature, it is a priori unclear how far this procedure can be pushed before reaching severe statistical limitations.
Accordingly, it is the aim of this work 
to propose 
a direct and 
convenient way to determine the $\lambda_k$'s, 
analogous to the one
available when the stationary state is
described by the canonical Gibbs ensemble.  In this case, the only 
unknown parameter $\beta$ 
can be determined by measuring dynamical quantities such as
fluctuation-dissipation relations (FDR's). 

FDR's establish a link between the 
time-delayed correlation and linear response functions: 
In Gibbs-Boltzmann equilibrium these FDR's have a universal form, which is independent of the specific 
system and observable under study inasmuch as they just involve the 
temperature $\beta^{-1}$ of the system. 
While they emerge quite naturally for classical and quantum systems in contact with an equilibrium environment 
\cite{Kubo-RMP,Cugliandolo-review}, 
the FDR's follow also from the so-called eigenvalue thermalization hypothesis~\cite{Deutsch91,Srednicki94,Khatami13} 
for isolated quantum systems.  
Out of canonical equilibrium, instead, 
these relations can actually be used in order to  
quantify the departure from equilibrium. One way to do this --- which has been pioneered in studies of complex systems in contact with thermal baths --- 
is to replace the equilibrium temperature by 
time- or frequency-dependent parameters~\cite{Cugliandolo97,Cugliandolo-review,CalabreseGambassi} which
play the role of non-equilibrium effective temperatures.   

In this work 
we show that the Lagrange multipliers $\lambda_k$'s of the GGE of a number of isolated (non-interacting) integrable systems 
which reach a stationary state 
can be read from the FDR's of properly chosen observables. 
We first present a generic argument supporting this statement, which is then illustrated with a number of examples:
a change of the mass of a bosonic field theory
and  quenches of the superlattice potential of hard-core bosons in one dimension, 
of the interaction in the Lieb-Liniger model, and of
the transverse field in the quantum Ising chain. For the latter, we briefly recall and extend published
results, that should be relevant to experimental checks.


\section{GGE and FDR's}
\label{sec:GGE-FDR}

In order to present our approach for determining the parameters which define the generalized Gibbs ensemble (GGE), we first recall  the definition of the GGE in Sec.~\ref{subsec:GGE} and then discuss how they can be inferred from the fluctuation-dissipation relations (FDR's) in Sec.~\ref{subsec:FDRs}.

\subsection{The generalized Gibbs ensemble}
\label{subsec:GGE}

Consider a generic integrable system with Hamiltonian $\hat H$
and a set of conserved charges  $\{ \hat{Q}_k \}$ which commute with $\hat H$.
The GGE density matrix $\hat{\rho}_{\rm GGE}$ is obtained by maximizing
the von Neumann entropy $S[\hat\rho] = \hat \rho \ln \hat \rho$ under the constraints that the 
expectation value $\langle \hat Q_k \rangle_{\rm GGE} =
\mbox{Tr} ( \hat Q_k \hat \rho_{\rm GGE} )$ of each charge $\hat Q_k$ on this GGE equals the corresponding value 
$\langle  \psi_0 |  \hat{Q}_k | \psi_0 \rangle$ taken on the initial state $| \psi_0 \rangle$, which is conserved by the evolution. 
Under these conditions  
$\hat{\rho}_{\rm GGE}$ takes the form
\begin{equation}
\label{density_matrix}
\hat{\rho}_{\rm GGE} = 
Z^{-1}_{\rm GGE} \; 
e^{
 - \sum_k \lambda_k  \hat{Q}_k }, 
\end{equation}
where $\lambda_k$ are the Lagrange 
multipliers enforcing the constraints 
and $Z_{\rm GGE}$ is a normalization factor.



\subsection{The fluctuation-dissipation relations}
\label{subsec:FDRs}

Take a system in a stationary state described by a generic density matrix $\hat \rho$
and set the units such that $\hbar =1$.
The time-dependent correlation $C$ and linear response function $R$
of a generic observable $\hat A$ can be expressed as \cite{Kubo-RMP}
\begin{eqnarray}
C(t_2,t_1) &=& \frac12 \langle [ \hat A(t_2), \hat A^\dag(t_1) ]_+ \rangle , 
\label{eq:def-C}
\\
R(t_2,t_1) &=& i \langle [ \hat A(t_2), \hat A^\dag(t_1) ]_- \rangle  \ \theta(t_2-t_1), 
\label{eq:def-R}
\end{eqnarray}
where $[\hat X,\hat Y]_\pm \equiv \hat X \hat Y\pm \hat Y \hat X$,
the expectation value $\langle \cdots \rangle$ is calculated over $\hat \rho$, while $\theta(t) =0$ for $t<0$ and 1 otherwise. 
By using the Lehmann representation, both $C$ and $R$ can be expressed as sums over a complete basis of eigenstates $\{ | n\rangle \}_{n\ge 0}$ of the Hamiltonian $\hat H$, with 
increasing eigenvalues $E_n$. Taking the 
Fourier transform  of Eqs.~\eqref{eq:def-C} and \eqref{eq:def-R} with respect to  $t_2-t_1$ one finds
\begin{eqnarray}
\label{FT_C}
C(\omega) \! = \!
\pi \!\!\! 
\sum_{m,n \ge 0} \!\!\!
\delta(\omega + E_n - E_m)  |A_{nm}|^2 
(\rho_{nn} + \rho_{mm}), \;\;\; &
\\
\label{FT_R}
\text{Im}\, R(\omega) \! = \!
 \pi \!\!\!  
\sum_{m,n \ge 0}  \!\!\!
\delta(\omega + E_n - E_m)  |A_{nm}|^2 
(\rho_{nn} - \rho_{mm}),\;\;\;&
\end{eqnarray}
where $A_{mn} = \langle m | \hat A | n \rangle$ and $\rho_{nn}=\langle n | \hat \rho | n \rangle $. The FDR in the 
frequency domain is the ratio between these two quantities which, in full generality, can be parametrized in terms of a
frequency- and observable-dependent effective temperature $\beta_{\rm eff}^{-1}$ as
\begin{equation}
\text{Im}\, R(\omega)/C(\omega) = 
\tanh\left(\beta_{\rm eff}(\omega)\omega/2\right). 
\end{equation}
For fermionic $\hat A$'s, $C(\omega)$ and 
$\text{Im}\, R(\omega)$ are exchanged in the equation. Clearly, these two functions do not vanish only if $\omega$ takes values within the discrete set  
$\{E_m - E_n\}_{m,n\ge 0}$ (due to the delta functions in Eqs.~\eqref{FT_C} and \eqref{FT_R} and  
the corresponding matrix element $A_{nm}$ of the operator $\hat A$ does not vanish.

In the case of Gibbs-Boltzmann equilibrium there is a single charge, $\rho_{nn} \propto \exp(-\beta E_n)$, and one immediately realizes that, 
for any bosonic $\hat A$, 
Eqs.~\eqref{FT_C} and \eqref{FT_R}  imply the celebrated fluctuation-dissipation theorem
\begin{equation}
 \text{Im} \, R(\omega) = \tanh(\beta\omega/2) \; C(\omega),
 \label{eq:FDT-1}
\end{equation}
for all frequencies $\omega$. Thus,  this equation
allows us to read the inverse temperature $\beta$ by simply taking the ratio between $\text{Im} \, R(\omega)$  and $C(\omega)$. 

In contrast, for the GGE, $\rho_{nn} \propto \exp(-\sum_k \lambda_k Q_{kn})$ with $Q_{kn}=\langle n | \hat Q_k |n\rangle$. 
The presence of several charges makes the relationship between $C(\omega)$ and $\text{Im}\, R(\omega)$ for generic observables
much more complicate than Eq.~\eqref{eq:FDT-1}. 
However,  by properly choosing the observable $\hat A$ we can
extract the $\lambda_k$'s from the corresponding FDR.
Let us show how this works, starting from a non-interacting Hamiltonian in its diagonal form
\begin{equation}\label{Hdiagonal}
\hat{H} = \sum_k \epsilon_k \, \hat{\eta}_k^{\dag}\hat{\eta}_k,
\end{equation}
where the $\hat{\eta}_k$'s are creation operators for bosonic or fermionic
excitations of energy $\epsilon_k$, labeled by  a set of quantum numbers $\{ k\}$. 
The operators $\hat{\eta}_k$ satisfy canonical commutation or anti-commutation relations.
The mutually commuting number operators   $\hat{Q}_k= \hat{\eta}_k^{\dag}\hat{\eta}_k$,
are the conserved charges. The GGE density matrix is
given by Eq.~(\ref{density_matrix}) with $\hat{Q}_k = \hat{\eta}_k^{\dag}\hat{\eta}_k$, while $\beta_k \equiv \lambda_k/\epsilon_k$ 
defines a mode-dependent inverse ``effective temperature''.
We now take an operator $\hat A$ of the form
\begin{equation}\label{Ex_A}
\hat A = \sum_{k} \ ( \alpha_k  \hat \eta_k + \alpha_k^\ast \hat \eta_k^\dag ),
\end{equation}
where $\alpha_k$ are some complex coefficients.
By plugging this $\hat A$ in Eqs.~(\ref{FT_C}) and (\ref{FT_R}) and taking their ratio 
for  $\omega  =   \omega_k  \equiv \epsilon_k $, in the absence of degeneracies with respect to $k$,
we obtain:
\begin{eqnarray}
\frac{ \text{Im} \, R(\omega_k) }{ C(\omega_k)  }
=   \tanh\Big( \frac{\lambda_k}{2} \Big) = \tanh\Big( \frac{\beta_k \epsilon_k}{2} \Big).
\end{eqnarray}
Accordingly, the $\beta_k$'s or, alternatively the $\lambda_k$'s, can be extracted 
from the ratio $\text{Im}\, R(\omega)/C(\omega)$, independently of
the value of the $\alpha_k$'s or of the other Lagrange multipliers, 
by simply choosing an adequate observable $\hat A$ and the 
frequency $\omega$ which selects a certain mode.
Note that one is not restricted to operators of the 
form~\eqref{Ex_A}, as the observable 
could be a many-body operator, as long as it involves a simple sum over 
one $k$-mode, such as, e.g., $\hat A = \sum_{k} \alpha_k {\hat \eta}^{\dag}_{k} {\hat \eta}^{\dag}_{k+k_1} \hat \eta_{k+k_2} \hat \eta_{k+k_3}$, with fixed $k_{1,2,3}$.
In addition, one should also require that 
the energy differences $E_n-E_m$ between each possible pair of many-body eigenstates with non-vanishing matrix elements 
$A_{nm}$ (see Eqs.~\eqref{FT_C} and \eqref{FT_R}) 
are not degenerate, apart from trivial 
symmetries such as those related to spatial inversion. 
%



\section{Four examples}

We show in this Section 
how to use the general strategy outlined in Sec.~\ref{sec:GGE-FDR} to determine the $\lambda_k$'s of four rather simple, but experimentally relevant, integrable models.

\subsection{The bosonic free field}

The bosonic free field theory is defined by
\begin{eqnarray}
\displaystyle \hat H =  \frac12 \int {\rm d}^d x \ \left[ \hat\pi^2(x) +  ( \partial_x\hat \phi(x))^2 + m^2 \hat \phi^2(x) \right],
\end{eqnarray}
where $\hat\phi_x$ and $\hat \pi_x$ are canonically conjugated
operators and $m$ is the mass. For simplicity, 
we consider here the case of one spatial dimension $d=1$ in the
thermodynamic limit, as the generalization to $d>1$ and finite volume is straightforward.
We consider the effect of an instantaneous change in the 
mass \cite{Pasquale,Spyros,Mitra} by preparing the system in the ground state with mass $m_0$, and later evolving it with a different mass $m$.
This Hamiltonian can be diagonalized by introducing bosonic creation and annihilation
operators $\{\hat b^\dag_k,\hat b_k\}$ with a dispersion relation $\epsilon_k(m)=\sqrt{m^2+k^2}$. 
The conserved quantities are $\hat Q_k \equiv \hat b_k^\dag \hat b_k $ with
\begin{equation}
n_k \equiv \langle \hat Q_k \rangle = 
(\epsilon_k-\epsilon_k^0)^2/(4 \epsilon_k \epsilon_k^0) \; ,
\end{equation}
where $\epsilon_k=\epsilon_k(m)$ and $\epsilon_k^0=\epsilon_k(m_0)$, while
the associated Lagrange multipliers in the GGE read~\cite{Pasquale}
\begin{equation}
\displaystyle 
\lambda_k =   2  \ln \left|
(\epsilon_k+\epsilon_k^0)/(\epsilon_k-\epsilon_k^0)\right|.
\label{eq:lambda-HO}
\end{equation}
A convenient observable to extract the $\lambda_k$'s from 
 an FDR turns out to be $\hat A = \hat \phi_k$.
The stationary contributions to its correlation and response functions (the non-stationary ones vanish upon averaging 
over the earliest time) are 
\begin{eqnarray}\label{Eq_CR_hcb}
\displaystyle C(t) &=& 
(4 \epsilon_k^2 \epsilon_k^0)^{-1} \, (\epsilon_k^2 +{\epsilon_k^0}^2) \, \cos(\epsilon_k t), 
\\
\displaystyle R(t) &=& (2\epsilon_k)^{-1} \, \sin(\epsilon_k t) \ \theta(t),
\end{eqnarray} 
where $t=t_2-t_1$.
The FDR ratio at $\omega=\omega_k  \equiv \epsilon_k $ reads
\begin{equation}
\displaystyle \text{Im}\, R(\omega_k)/C(\omega_k) =
\tanh (\lambda_k/2), 
\end{equation} 
with $\lambda_k$ given  by Eq.~\eqref{eq:lambda-HO}. 
Using  the results in Ref.~\cite{Spyros}  one can easily verify that this equality
holds also for quenches from thermal initial states, with the 
corresponding
$\lambda_k$'s. 
Moreover, in the classical limit, i.e., in a system of independent and \emph{classical}  harmonic oscillators evolving after a quench of the oscillator's frequencies, 
there is a similar link between the FDR and the parameters describing the stationary probability distribution, not only for thermal but 
also for generic initial conditions. 


\subsection{Hard-core bosons in one spatial dimension}
\label{subsec:HCB}

Let us now turn to a one-dimensional 
system of hard-core bosons prepared in the ground state of
a superlattice potential of strength $\Delta$, later subject to a
quench~\cite{Rigol06,Chung12b,Bortolin15}. 
This problem maps onto a free-fermion Hamiltonian, 
\begin{eqnarray}
\displaystyle\hat H_0 =  \sum_{i=1}^L \; [ -\hat f^{\dag}_i \hat f_{i+1} - \hat f^{\dag}_{i+1} \hat f_i + \Delta  (-1)^i \hat f^{\dag}_i \hat f_i ] \; ,
\end{eqnarray}
where we assume periodic boundary conditions.
The ground state displays correlations, also referred to as
bipartite entanglement, since $\langle \hat f^\dag_{k+\pi} \hat f_k \rangle \neq 0$.
The quench consists in switching off the potential at time $t=0$, i.e., in setting $\Delta =0$ for $t>0$.
The conserved charges $\hat Q_k$ are the occupations of the fermionic modes, $ \hat Q_k =   \hat f^{\dag}_k \hat f_k $ for $ k \in [-\pi,\pi]$, with~\cite{Chung12b,Bortolin15}
\begin{equation}
\displaystyle n_k = \langle \hat Q_k \rangle 
= (1- \epsilon_k/\epsilon^0_k )/2 ,
\label{nk_FF}
\end{equation} 
where $\epsilon^0_k = \sqrt{4\cos^2 k+\Delta^2}$ and 
$\epsilon_k = -2\cos k$.
The expression of the  $\lambda_k$'s in terms of the $\epsilon_k$'s and $\epsilon_k^0$'s 
turns out to be formally identical to 1/2 of the r.h.s.~of Eq.~\eqref{eq:lambda-HO}. 
For $|\Delta | \gg | \epsilon_k|$, $\lambda_k \to 2\epsilon_k /|\Delta|$ and the GGE reduces to the 
Gibbs-Boltzmann form with temperature $|\Delta|/2$. 
A natural observable to extract the $\lambda_k$'s from the FDR is the density $\hat \rho(q,t)$ of bosons which, in Fourier space, reads
\begin{equation}\label{rho}
\displaystyle\hat \rho(q,t) =  \sum_{k} \ e^{- i (\epsilon_{k}-\epsilon_{k-q}) t} \ \hat f_{k-q}^{\dag} \hat f_{k}. 
\end{equation}
The corresponding correlation and response functions can be obtained from $\langle [ \hat  \rho(q,t_2), \hat \rho(-q,t_1) ]_{\pm} \rangle$ in the stationary limit:
\begin{eqnarray}\label{CR_hcb1}
\displaystyle C(q,t) 
\! &=&  \!  \sum_{k} e^{- i (\epsilon_{k}-\epsilon_{k-q}) |t|} 
\left(   \frac{n_{k-q} + n_{k}}{2} -   n_{k-q} n_{k} \right) \! ,
\;\;\;\; \\ \label{CR_hcb2}
R(q,t) \! &=& \! i \theta(t)  \sum_{k} e^{- i (\epsilon_{k}-\epsilon_{k-q}) t}
(  n_{k-q} - n_{k} ),
\end{eqnarray}
with $t=t_2-t_1$ and $q \neq 0,\pi$. 
After a Fourier transformation with respect to $t$, 
the frequency $\omega$ selects the values of $k$ in the sum such that
$\omega=\omega_{k,q}\equiv\epsilon_{k}-\epsilon_{k-q}$.
 For a given $q$, this condition is satisfied by two values
$k_1(q,\omega)$ and $k_2(q,\omega)$.  If we choose $\omega = 
\omega_q \equiv
2\epsilon_{(q+\pi)/2}$, these $k$'s are forced to coincide: $k_{1,2}=k_q \equiv 
(q+\pi)/2$, 
thus selecting a single mode in the sum.
The FDR then becomes a simple ratio between the factors in parentheses in the sums that, in 
turn, can be recast as functions of the $\lambda_q$'s for this model.  
Exploiting their explicit expressions, we finally obtain
\begin{equation}
\text{Im}\, R(q,\omega_q)/C(q,\omega_{q})
=  \tanh \lambda_{k_q}. 
\label{eq:single-lambdak}
\end{equation}


\subsection{The Lieb-Liniger model}
\label{subsec:LL}

A closely related example is provided by the quench of the Lieb-Liniger Hamiltonian~\cite{LiebLiniger63},
which describes the behaviour of bosons in $d=1$ with pairwise delta-interactions, 
\begin{equation}\label{H_LL}
\displaystyle \hat H \! = \! \int \! {\rm d}x  \ [  \partial_x \hat \phi^{\dag}(x)  \partial_x \hat \phi(x) + c \  \hat \phi^{\dag}(x) \hat  \phi^{\dag}(x) \hat \phi(x) \hat \phi(x) ],
\end{equation}
where $\hat \phi$ is the canonical bosonic field and $c$ is the coupling constant.
The system is prepared in the ground state of the non-interacting Hamiltonian with $c=0$
and a particle density $\varrho$: A quench of the interaction strength is then performed
such that
the subsequent dynamics occur with $c\to\infty$~\cite{Kormos14}. 
In this limit the Hamiltonian can be written in terms of hard-core bosons $\hat \Phi(x)$.
A Jordan-Wigner transformation maps the Hamiltonian 
of these bosons onto the one of free fermions $\{\hat f_k\}$ and, after a Fourier transform, it becomes:
$\hat H = \sum_{k=-\infty}^{\infty} \epsilon_k \hat f^{\dag}_k \hat f_k $,  
where $\epsilon_k=k^2$.
As in Eq.~(\ref{nk_FF}) the fermionic occupation numbers $n_k = \langle \hat f^{\dag}_k \hat f_k \rangle$  are conserved 
with~\cite{Kormos14}  
\begin{equation}
n_k =  4 \varrho^2/(\epsilon_k+4\varrho^2)  \quad \mbox{and} \quad \lambda_k = \ln( \epsilon_k/(4\varrho^2)).
\label{eq:nk-lambdak-Lieb-Liniger}
\end{equation}
The connected 
density-density correlation function was calculated in Ref.~\cite{Kormos14} and, 
in the stationary limit, it reads 
\begin{equation}
\begin{split}
&
\displaystyle \langle \hat \rho(q,t_2) \hat \rho(-q,t_1) \rangle 
\\ 
&
\qquad
 = \displaystyle \int \frac{{\rm d} k}{2\pi} \ e^{ - i ( \epsilon_{k} - \epsilon_{k-q} )(t_2-t_1)} \ n_{k-q} ( 1 - n_{k} ),
\end{split}
\end{equation}
from which we find that the symmetrized correlation and linear response functions are, in terms of the $n_k$'s,
formally identical to Eqs.~\eqref{CR_hcb1} and \eqref{CR_hcb2}, respectively. 
In this case, for a fixed value of $q$, each frequency $\omega$ selects a {\it single} mode $k$ such
that $\omega=\omega_{k,q} = \epsilon_{k}-\epsilon_{k-q} = 2kq-q^2$ and the FDR  takes the form 
\begin{equation}
\text{Im}\, R(q,\omega)/C(q,\omega)|_{\omega=\omega_{k,q}} = 
\tanh\left((\lambda_{k} - \lambda_{k-q})/2\right).
\label{eq:FDR-2lambdas}
\end{equation}  
The  $\lambda_k$'s --- given by Eq.~(\ref{eq:nk-lambdak-Lieb-Liniger}) --- can be extracted one by one from this formula. 
This is achieved by evaluating the ratio on its l.h.s.~(equal to $2q^2\omega/(q^4+\omega^2)$) for fixed $q$ at the particular frequency $\omega = \omega_q \equiv \pm q\sqrt{16 \varrho^2 + q^2}$,
which selects the mode $k_{q}
\equiv (q^2 + \omega_q)/(2q)$ and turns Eq.~(\ref{eq:FDR-2lambdas}) into Eq.~(\ref{eq:single-lambdak}).

\begin{figure}
\!\!\includegraphics[scale=0.101]{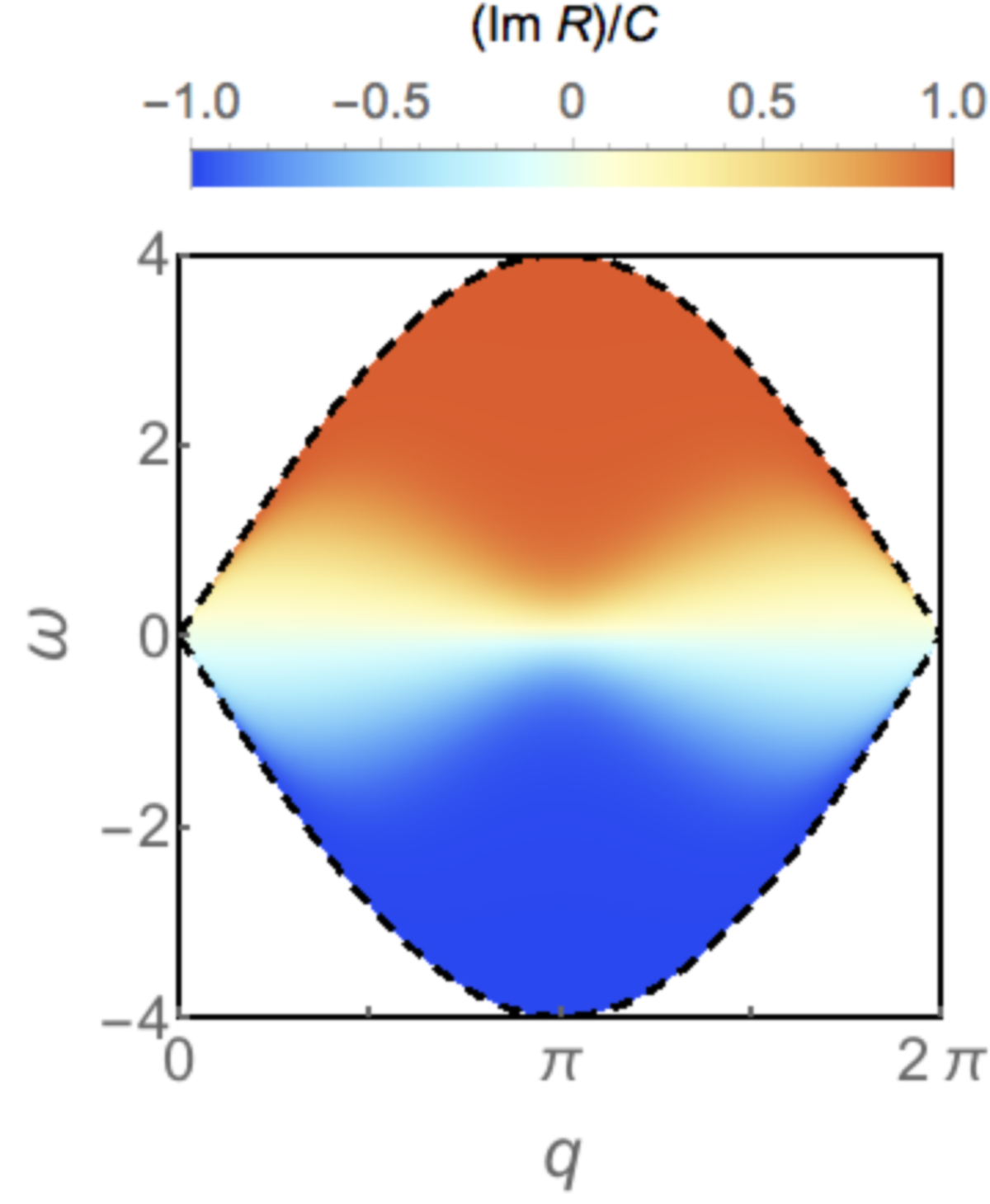}  
\includegraphics[scale=0.101]{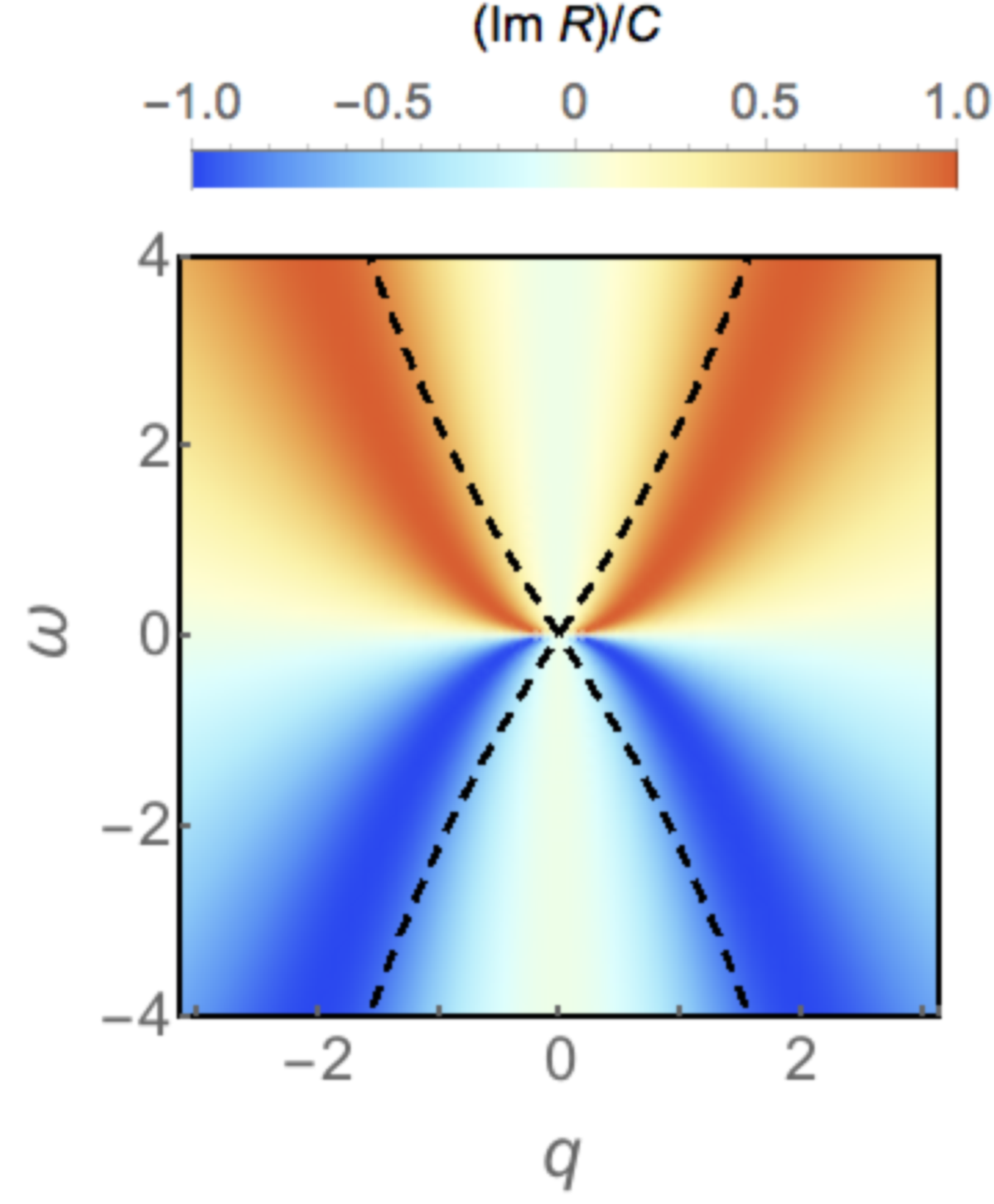}
\caption{(Color online) Contour plots of $\text{Im}\, R(q,\omega)/C(q,\omega)$ as a function of $q$ and $\omega$ for the quench of hard-core bosons in the superlattice potential with $\Delta=0.5$ (left) and for the Lieb-Liniger model (right). The dashed lines indicate the values of this ratio which are relevant for determining $\lambda_k$'s, as explained in the main text. Due to the presence of the lattice, the ratio on the left panel is defined only within the domain shown.}
\label{Fig:FDTratio}
\end{figure}

In Fig.~\ref{Fig:FDTratio} we show the plot of the ratio $\text{Im}\, R/C$ as a function of $q$ and $\omega$
for the quench of the periodic potential studied in Sec.~\ref{subsec:HCB} (left) and for the quench of the Lieb-Liniger model discussed here (right).
In both cases the dashed line indicates the values of $(q,\omega)$ for which Eq.~\eqref{eq:single-lambdak} holds.


\subsection{The quantum Ising chain in a transverse field}

These ideas can also be applied to magnetic systems.
Consider, for example, the quantum Ising spin chain in a transverse field, with periodic boundary conditions,
\begin{equation}
\displaystyle \hat H =  - \sum_{j=1}^L \left(  \hat \sigma_j^x  \hat \sigma_{j+1}^x + \Gamma \hat \sigma_j^z \right), 
\label{eq:H-Ising}
\end{equation}
where $\hat \sigma^{x,z}_j$ are the standard Pauli matrices, 
and prepare it in the ground state with $\Gamma = \Gamma_0$. 
The quench consists in a change $\Gamma_0\to\Gamma$ 
by a finite amount. After subsequent Jordan-Wigner and 
Bogolioubov trasformations the system becomes 
a set of non-interacting fermions 
with dispersion $\epsilon_k(\Gamma) = 2 \sqrt{1+\Gamma^2-2\Gamma\cos k}$.
The conserved quantities are the mode occupation numbers 
$n_k = \langle \hat Q_k \rangle = \langle \hat f_k^\dag\hat f_k \rangle$ as in Eq.~(\ref{nk_FF})
with $\epsilon_k = \epsilon_k(\Gamma)$, 
$\epsilon^0_k = \epsilon_k(\Gamma_0)$,
and~\cite{Karevski06,Dziarmaga10,Dutta10,Foini11,Foini12,Fagotti12}
\begin{equation}\label{lambda_ising}
\displaystyle \lambda_k =  2 \mbox{\;arctanh} \left(4  \frac{\Gamma\Gamma_0-(\Gamma+\Gamma_0)\cos k +1}{\epsilon_k(\Gamma)\epsilon_k(\Gamma_0)}\right). 
\end{equation}
A natural observable for extracting $\lambda_k$'s is the total transverse magnetization
$ \hat M^z_{q=0}(t)$, where
\begin{equation}
\displaystyle \hat M^z_q(t) = \frac{1}{L} \sum_{j=1}^L \ e^{i q j} \hat \sigma_j^z(t), 
\label{eq:Mz-Ising}
\end{equation}
that can be written as a bi-linear combination of the fermionic creation and annihilation 
operators $\hat f^\dag_k$ and $\hat f_k$. The FDR for (connected) correlations of this observable 
at $q=0$ was calculated in Refs.~\cite{Foini11,Foini12} (in particular, see Sec.~4.2 and Eq.~(73) in Ref.~\cite{Foini12}) 
and it was there recognized that the $\lambda_k$'s arise from this FDR in a similar way  to what we have already explained for other models. 
Another operator accessible in the laboratory 
is the Fourier transform of the two-time correlation and response functions
of the transverse spin
$\hat \sigma^z_j(t)$, that can be expressed in terms of 
$[\hat M^z_q(t_2),\hat M^z_{-q}(t_1)]_{\pm}$. 
It is then natural to ask whether the $\lambda_k$'s in Eq.~\eqref{lambda_ising} measured using $q=0$ are 
compatible with the ones that one can extract from measurements at $q\neq 0$. 
This is indeed the case. For example, for $-q_{\rm max} \leq q \leq q_{\rm max}$ with $q_{\rm max} = 2 \arccos \rm{min}\{\Gamma,\Gamma^{-1}\}$ 
one can check that the $\lambda_k$'s satisfy
\begin{eqnarray}
\text{Im} \, R(q, \omega_{k,q})/C(q,\omega_{k,q}) = \tanh\left((\lambda_k + \lambda_{k+q})/2\right),
\label{eq:lambda-mq}
\end{eqnarray}
where the measuring $\omega$ selects the $k$ such that  $\omega=\omega_{k,q}\equiv \epsilon_k+\epsilon_{k+q}$, 
as detailed in Appendix~\ref{app:Ising}.

If the chain is prepared, instead, in a pre-quench state in equilibrium  at temperature $\beta_0^{-1}$, the corresponding $\lambda_k$'s are given by Eq.~\eqref{lambda_ising} in which the term in square brackets is multiplied by $1 - 2 n_k^0$, where $n_k^0 = (1+e^{\beta_0\epsilon_k^0})^{-1}$ is the pre-quench population. A direct but 
long calculation shows that also these $\lambda_k(\beta_0)$'s are  recovered from 
Eq.~(\ref{eq:lambda-mq}).

\section{Conclusions}

We have explained how the {\it complete set} of 
Lagrange multipliers $\lambda_k$'s of a GGE can be obtained at once from a {\it single} 
measurement of time-delayed correlation and linear response functions of suitable physical 
observables, that are within experimental reach. 
We focused our analysis on the steady state of integrable systems but our arguments are 
sufficiently general to be also applicable to quasi-stationary prethermal states described by a GGE. 
Although we discussed primarily quenches from ground states, the proposed approach is  viable also for generic initial states, including thermal ones. 
In practice, having in mind the experimental setting of Ref.~\cite{GGE-exp}, one should measure 
the space-time dependent correlation and response functions of the bosonic density and calculate their
Fourier transforms in both space and time. 
By sweeping the corresponding momenta and frequencies in the manner described above for, e.g., the Lieb-Liniger model, 
one then extracts all $\lambda_k$'s.

We are currently working on the extension of these ideas to {\it
  interacting} systems, such as quenches to generic $c$ in the
integrable Lieb-Liniger model in Eq.~\eqref{H_LL} 
 --- using methods employed in
Ref.~\cite{vandenBerg,Jacopo} or the quench action approach \cite{QAreview} --- or in the solvable 
$O(N)$ model for $N\to\infty$, which features a variety of dynamical 
phases with different properties but has no obvious GGE description, see, 
e.g., Refs.~\cite{Chandran,Chiocchetta}.

\smallskip 

\begin{acknowledgments}
We thank M.~Rigol and T.~Giamarchi for useful discussions. LFC and RK
initiated this work while visiting the KITP, an institute supported in
part by the National Science Foundation under Grant No. NSF
PHY11-25915.
LFC, LF and RK also would like to thank SISSA for 
hospitality.  RK's research effort here was supported primarily by the
U.S. Department of Energy (DOE) Division of Materials Science under
Contract No. DE-AC02-98CH10886.
LFC is a member of the Institut Universitaire de France. 
This work was supported in part by the Swiss SNF under Division II and by the ERC under the Starting Grant 279391 EDEQS.
\end{acknowledgments}

\appendix

\section{Quench of the Ising spin chain}
\label{app:Ising}

This Appendix provides some details on the solution of the analysis of the quantum Ising spin chain.
We recall in Sec.~\ref{sec:Ising-quench} the solution of the post-quench dynamics of this model.
In Sec.~\ref{sec:FDR-Ising} we calculate  the correlation and response functions of the transverse magnetization,
and we show how the Lagrange multipliers $\lambda_k$ of the GGE of  can be determined on the basis of their FDR.

\subsection{Post-quench dynamics}
\label{sec:Ising-quench}

The Hamiltonian of the quantum Ising spin chain, reported in Eq.~\eqref{eq:H-Ising}, 
can be diagonalized by means of
a Jordan-Wigner transformation and a Fourier transform which maps
spins into fermions $\{\hat c_k, \hat c_{-k}^\dagger\}_k$. 
The dynamics of these fermions after a quench of the transverse field $\Gamma_0\to\Gamma$ 
can be exactly determined in terms of the quasi-particles $\{\hat{\gamma}_k^{\Gamma_0}, \hat{\gamma}_{-k}^{\Gamma_0\,\dag}\}_k$ which diagonalize the pre-quench Hamiltonian $\hat H(\Gamma_0)$, as explained, e.g., in Ref.~\cite{Foini12}, which we refer to for additional details as well as  for the notation adopted here.  In particular, the evolution of the fermions reads: 
\beq
\displaystyle \left( \begin{array}{c} 
 \hat{c}_k (t)  \\[2mm]
\hat{c}_{-k}^{\dag}(t)  \end{array} \right) = 
\left( \begin{array}{cc}
u_k^{\Gamma,\Gamma_0}(t) & - [v_k^{\Gamma,\Gamma_0}(t)]^*  \\[2mm]
 v_k^{\Gamma,\Gamma_0}(t) & [u_k^{\Gamma,\Gamma_0}(t)]^*
 \end{array} \right)
\left( \begin{array}{c}
\hat{\gamma}_k^{\Gamma_0}  \\[2mm]
\hat{\gamma}_{-k}^{\Gamma_0 \; \dag}  \end{array} \right),
\label{Eq:Dynamics_quench}
\eeq
where
\begin{widetext}
\beq
\begin{cases}
\displaystyle u_k^{\Gamma,\Gamma_0}(t) = e^{-i \epsilon^\G_k t}   \cos \theta_k^{\Gamma} \cos( \theta_k^{\Gamma}- \theta_k^{\Gamma_0})
+ e^{i  \epsilon^\G_k t}  \sin \theta_k^{\Gamma} \sin( \theta_k^{\Gamma}- \theta_k^{\Gamma_0}) ,
\\
\displaystyle v_k^{\Gamma,\Gamma_0}(t) = i e^{- i \epsilon^\G_k t}\sin \theta_k^{\Gamma} \cos( \theta_k^{\Gamma}- \theta_k^{\Gamma_0})
-  i e^{i  \epsilon^\G_k t} \cos \theta_k^{\Gamma} \sin( \theta_k^{\Gamma}- \theta_k^{\Gamma_0}) ,
\end{cases}
\label{eq:ukvk}
\eeq
\end{widetext}
with $\epsilon^\G_k \equiv \epsilon_k(\Gamma)=2\sqrt{1+\Gamma^2-2\Gamma\cos k}$, while the so-called Bogoliubov angles $\theta_k^{\Gamma,\Gamma_0}$ are determined by solving 
$\tan (2 \theta_{k}^{\Gamma}) = (\sin k)/(\Gamma-\cos k)$,
with the important property that $\theta_{-k}^{\Gamma} = - \theta_k^{\Gamma}$, which turns out to be useful later on.
The initial state of the evolution, i.e., the ground state of the pre-quench Hamiltonian $\hat H(\Gamma_0)$ is actually the vacuum of the fermionic quasi-particles  $\{\hat{\gamma}_k^{\Gamma_0}, \hat{\gamma}_{-k}^{\Gamma_0\,\dag}\}_k$.
For convenience, below we denote by  
$\Delta_k \equiv \theta_k^{\Gamma}- \theta_k^{\Gamma_0}$ the difference between the angles appearing in Eq.~\eqref{eq:ukvk}, 
which however differs from the notation $\Delta_k \equiv 2(\theta_k^{\Gamma}- \theta_k^{\Gamma_0})$ used 
in Ref.~\cite{Foini12}.

\subsection{FDR for the transverse magnetization}
\label{sec:FDR-Ising}

\begin{figure}[h!!]
\!\!\includegraphics[width=0.3\textwidth]{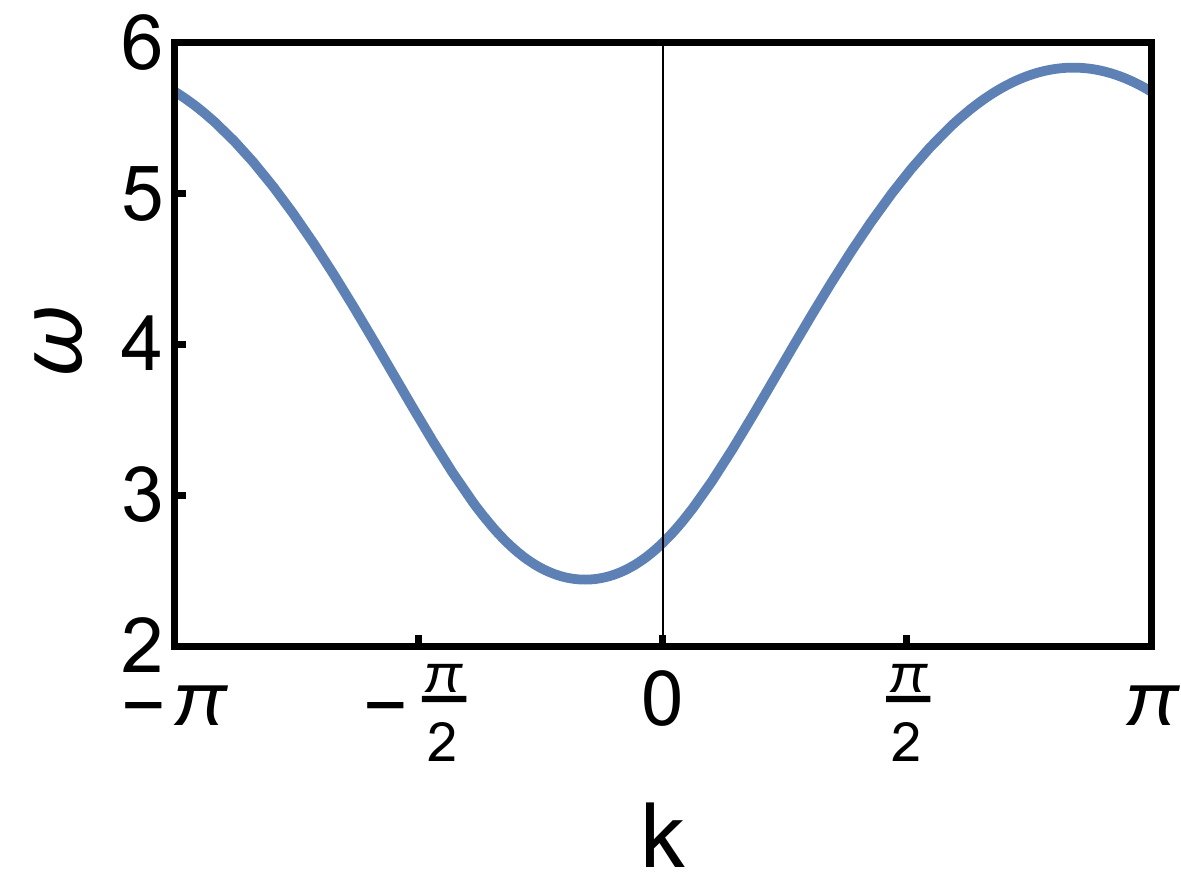}  
\!\!\includegraphics[width=0.3\textwidth]{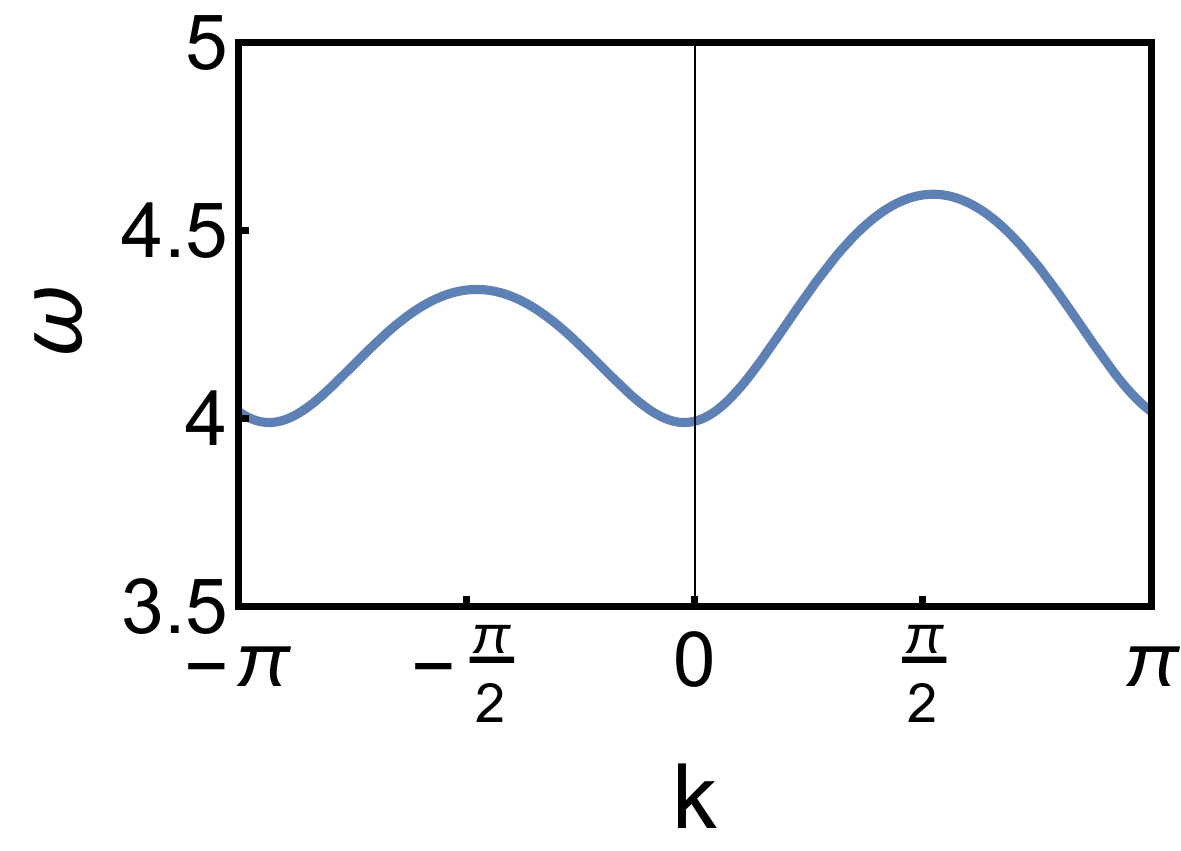}  
\caption{(Color online) Plot of the dependence of $\omega=\epsilon_k+\epsilon_{k+q}$ on $k$
 for $q=1$ (left) and $q=3$ (right), with $\Gamma =1/2$. For sufficiently large values of $q$, i.e., $|q| > q_{\rm max}$ (see the main text) there is a wide range of values of $\omega$ for which more than two values of $k$ satisfy Eq.~\eqref{eq:cond-S}.
}\label{Fig:omega}
\end{figure}

According to the definition of $\hat M^z_q$ in Eq.~\eqref{eq:Mz-Ising}, 
we consider the quantity,
%
\beq
\begin{split}
&\langle \hat M^z_q(t_1) \hat  M^z_{-q}(t_2)\rangle_c  = \\
&= \langle \hat  M^z_q(t_1) \hat  M^z_{-q}(t_2)\rangle - \langle \hat  M^z_q(t_1)\rangle  \langle \hat  M^z_{-q}(t_2)\rangle\\ 
& =  4 \sum_{k_1,k_2} \ 
 \langle \hat c^{\dag}_{k_1+q}(t_1) \hat c_{k_1}(t_1) \hat c^{\dag}_{k_2}(t_2) \hat c_{k_2+q}(t_2) \rangle,
 \end{split}
\eeq
where we used the fact that $\hat \sigma^z_j = 1 - 2 \hat c^\dag_j \hat c_j$ in terms of the fermions, see, e.g., Ref.~\cite{Foini12}.
Substituting the explicit expressions for the evolution of the fermionic operators provided by Eq.~\eqref{Eq:Dynamics_quench} and taking the expectation value on the pre-quench ground state, one obtains (for $q\neq 0$, with additional contributions for $q=0$) 
\beq
\langle \hat M^z_q(t_1) \hat M^z_{-q}(t_2)\rangle_c =  4 (A + B),
\label{eq:S-corr}
\eeq 
with
\beq
A \equiv  \sum_{k} \ v_{k+q}(t_1)    u_{k}(t_1) u_{k}^\ast(t_2) v_{k+q}^{\ast}(t_2) 
\eeq
and 
\beq
B \equiv \sum_{k} \ v_{k+q}(t_1)    u_{k}(t_1)  u_{k+q}^\ast(t_2) v_{k}^{\ast}(t_2).
\eeq
%
In order to simplify the notation, we omitted the superscripts $\Gamma, \Gamma_0$ from Eq.~\eqref{Eq:Dynamics_quench}.
We are interested in the stationary limit of the correlation function in Eq.~\eqref{eq:S-corr}.
Keeping solely the terms that depend only on $t \equiv t_2-t_1$,
one obtains
\begin{widetext}
\beq
\begin{split}
\displaystyle A =    \sum_k   & \Big[  
e^{i (\epsilon_{k+q}+\epsilon_k)t} \left( \sin \theta_{k+q}\cos\Delta_{k+q}\cos\theta_k \cos\Delta_k \right)^2
+ e^{i (\epsilon_{k+q}-\epsilon_k)t} \left( \sin \theta_{k+q}\cos\Delta_{k+q}\sin\theta_k \sin\Delta_k \right)^2 \\
&+ e^{-i (\epsilon_{k+q}-\epsilon_k)t} \left( \cos \theta_{k+q}\sin\Delta_{k+q}\cos\theta_k \cos\Delta_k \right)^2
+ e^{-i (\epsilon_{k+q}+\epsilon_k)t} \left( \cos \theta_{k+q}\sin\Delta_{k+q}\sin\theta_k \sin\Delta_k \right)^2
\Big]
 \end{split}
\eeq
\beq
\begin{split}
\displaystyle B = &  \displaystyle  \sum_k \Big[  
e^{i (\epsilon_{k+q}+\epsilon_k)t}  \sin \theta_{k}\cos^2\Delta_{k} \cos\theta_{k+q} \sin\theta_{k+q} \cos^2\Delta_{k+q} \cos\theta_k
\\ & \displaystyle \qquad 
- e^{-i (\epsilon_{k+q}-\epsilon_k)t}  \sin \theta_{k}\cos^2\Delta_{k} \cos\theta_{k+q} \sin\theta_{k+q} \sin^2\Delta_{k+q} \cos\theta_k
 \\ %
& \displaystyle \qquad - e^{i (\epsilon_{k+q}-\epsilon_k)t}  \cos \theta_{k}\sin^2\Delta_{k} \cos\theta_{k+q} \sin\theta_{k+q} \cos^2\Delta_{k+q} \sin\theta_k
\\ & \displaystyle  \qquad 
+ e^{-i (\epsilon_{k+q}+\epsilon_k)t}  \cos \theta_{k}\sin^2\Delta_{k} \cos\theta_{k+q} \sin\theta_{k+q} \sin^2\Delta_{k+q} \sin\theta_k
\Big],
 \end{split}
\eeq
\end{widetext}
where, for convenience, we simplified the notation by omitting the superscript $\Gamma$ in $\epsilon^\Gamma_k$,  as no confusion can arise. Accordingly, the stationary part of Eq.~\eqref{eq:S-corr} is given by
$\langle \hat M^z_q(t_1) \hat M^z_{-q}(t_2)\rangle_c^{\text{stat}} =   4 \sum_k \left[ A_{12}^{(1)} + B_{12}^{(1)} + B_{12}^{(2)} + A_{12}^{(2)}\right]$ where
\begin{widetext}
\beq
\begin{split}
&A_{12}^{(1)} =  
 e^{i (\epsilon_{k+q}+\epsilon_k)t } \cos^2\Delta_{k} \sin\theta_{k+q} \cos^2\Delta_{k+q} \cos\theta_k \sin(\theta_k+\theta_{k+q})
\\ %
&B_{12}^{(1)} =  e^{- i (\epsilon_{k+q}-\epsilon_k) t} \cos^2\Delta_{k} \cos\theta_{k+q} \sin^2\Delta_{k+q} \cos\theta_k \cos(\theta_k+\theta_{k+q})
\\ %
&B_{12}^{(2)} =  \displaystyle e^{i (\epsilon_{k+q}-\epsilon_k) t}  \sin^2\Delta_{k}  \sin\theta_{k+q} \cos^2\Delta_{k+q} \sin\theta_k ( - \cos(\theta_{k+q}+\theta_k) )
\\ %
&A_{12}^{(2)} =  \displaystyle e^{-i (\epsilon_{k+q}+\epsilon_k) t}  \sin^2\Delta_{k}  \cos\theta_{k+q} \sin^2\Delta_{k+q} \sin\theta_k 
\sin(\theta_{k+q}+\theta_k) .
 \end{split}
\eeq
\end{widetext}
The correlation and response functions $C$ and $R$ of the magnetization $\hat M^z_q$ are defined, as usual, using the commutator and anti-commutator of $\hat M^z_q(t_1)$ and $\hat M^z_{-q}(t_2)$ [see Eqs.~\eqref{eq:def-C} and \eqref{eq:def-R}], 
which can be determined from the previous expressions. In particular, considering the commutator and anti-commutator $ \langle \hat M^z_q(t_1) \hat M^z_{-q}(t_2)\rangle_c^{\text{stat}} \mp  \langle \hat M^z_{-q}(t_2) \hat M^z_{q}(t_1)\rangle_c^{\text{stat}}$ one has, respectively:
 %
\begin{widetext}
\beq
\begin{split}
A_{12}^{(1)} \mp A_{21}^{(1)} + A_{12}^{(2)} \mp A_{21}^{(2)} &=
%
\sin(\theta_k+\theta_{k+q}) \left( \cos^2\Delta_{k} \cos^2\Delta_{k+q}  \mp  \sin^2\Delta_{k} \sin^2\Delta_{k+q}  \right)
\\
& 
 \qquad\times
 \left[ e^{i(\epsilon_k+\epsilon_{k+q}) t}  \sin\theta_{k+q} \cos\theta_k  \mp e^{- i(\epsilon_k+\epsilon_{k+q}) t}   \cos\theta_{k+q} \sin\theta_k  \right]
 \end{split}
\eeq
\beq
\begin{split}
B_{12}^{(1)} \mp B_{21}^{(1)} + B_{12}^{(2)} \mp B_{21}^{(2)} & = 
 \cos(\theta_k+\theta_{k+q}) \left( \cos^2\Delta_{k} \sin^2\Delta_{k+q}  \mp  \sin^2\Delta_{k} \cos^2\Delta_{k+q}   \right)
 \\
 &
\qquad\qquad  \times
 \left[ e^{i(\epsilon_k-\epsilon_{k+q}) t}  \cos\theta_{k+q}
 \cos\theta_k  - e^{- i(\epsilon_k-\epsilon_{k+q}) t}
 \sin\theta_{k+q} \sin\theta_k  \right],
 \end{split}
\eeq
\end{widetext}
where $A^{(i)}_{21}$ and $B^{(i)}_{21}$ are obtained from the corresponding $A^{(i)}_{12}$ and $B^{(i)}_{12}$ by replacing $t$ with $-t$ and by exchanging the subscripts $k+q$ and $k$ in the previous expressions.
Taking the Fourier transform in time of the stationary parts of $C(t_2,t_1)$ and $R(t_2,t_1)$ one obtains the corresponding two expressions $C(q,\omega)$ and $R(q,\omega)$ which depend also on the specific value of $q$ and  involve a sum over $k$ of a linear combination of $\delta (\omega \pm (\epsilon_k + \epsilon_{k+q}))$ and $\delta(\omega \pm (\epsilon_k - \epsilon_{k+q}))$ resulting from the Fourier transform of the exponentials in the previous expressions. We first note that if $\omega>0$ and $q$ are chosen in such a way that $\text{Im}\, R(q,\omega)$ and $C(q,\omega)$ receive contributions from $\delta(\omega - (\epsilon_k + \epsilon_{k+q}))$, then the remaining $\delta$'s do not contribute, as $\epsilon_k + \epsilon_{k+q} > |\epsilon_k - \epsilon_{k+q}|$ for $k\neq 0$.
On the other hand, for a given $\omega$ and $q$, if a certain value $k'$ of the index $k$ of the sum contributes to $C$ and $R$ because $\omega = \epsilon_{k'} + \epsilon_{k'+q}$, then also $-k'-q + 2\pi n$ (with integer $n$) does.  This means that, generically, the condition: 
\begin{equation}
\omega = \omega_{k,q} \equiv \epsilon_{k} + \epsilon_{k+q}
\label{eq:cond-S}
\end{equation}
is satisfied (depending on $q$ and $\omega$) by either none or an even number of values of $k\in [0,2\pi]$, with at most four values, as a direct inspection of Eq.~\eqref{eq:cond-S} as well as Fig.~\ref{Fig:omega} show.
%
%
%
%
%
In particular, from a more careful analysis of Eq.~\eqref{eq:cond-S}, one concludes that for $|q| < q_{\rm max}$ with $q_{\rm max} = 2 \arccos \rm{min}\{\Gamma,\Gamma^{-1}\}$, the equation has at most two solutions which are related as discussed above. Based on this relationship and on the symmetry properties of the Bogoliubov angles, one immediately concludes that these two selected values of $k$ results in the same contribution both to $C$ and to $\text{Im}\, R$ and 
therefore
\begin{widetext}
\begin{equation}
\begin{split}
\frac{\text{Im}\, R(q,\omega_{k,q})}{C(q,\omega_{k,q})} &= \frac{\sin(\theta_{k}+\theta_{{k}+q})
\left(\sin \theta_{{k}+q}\cos\theta_{k}+\sin\theta_{k}\cos\theta_{k+q}\right)
\left(\cos^2\Delta_{{k}+q}\cos^2\Delta_{k} - \sin^2\Delta_{{k}+q}\sin^2\Delta_{k}\right) }{\sin(\theta_{k}+\theta_{{k}+q})
\left(\sin\theta_{k+q}\cos\theta_{k}+\sin\theta_{k}\cos\theta_{k+q}\right)\left(\cos^2\Delta_{{k}+q}\cos^2\Delta_{k} + \sin^2\Delta_{{k}+q}\sin^2\Delta_{k}\right)}\\
&=
\frac{\cos^2\Delta_{{k}+q}\cos^2\Delta_{k} - \sin^2\Delta_{{k}+q}\sin^2\Delta_{k}}{\cos^2\Delta_{{k}+q}\cos^2\Delta_{k} + \sin^2\Delta_{{k}+q}\sin^2\Delta_{k}} \\
&=
\displaystyle  \frac{\cos(2\Delta_{k+q})+\cos (2 \Delta_k)}{1+\cos(2\Delta_{k+q})\cos (2 \Delta_k)} = \tanh\left(\frac{\lambda_k+\lambda_{k+q}}{2}\right),
\end{split}
\end{equation}
\end{widetext}
where in the last step we have used the fact that $\cos(2\Delta_k)=\tanh(\lambda_k/2)$ \cite{Foini12}. In particular, taking into account that $\lambda_k = \lambda_{-k}$, one can use the previous relationship in order to extract some of the $\lambda_k$'s, e.g., by choosing  $q=-2k$, with $\omega = \omega_{k,-2k} = 2 \epsilon_k$ and $|k| < q_{\rm max}/2$.
For $|q|>q_{\rm max}$, instead, Eq.~\eqref{eq:cond-S} can have more than two solutions (see Fig.~\ref{Fig:omega}) which are not all related as discussed above and therefore the corresponding contributions to the correlation and response functions are not necessarily equal and the resulting ratio $\text{Im}\, R/C$ depends in general also on $\theta_k$.

\end{document}